# Coherent 1-μm-Band Co-Transmission with C+L-Band Channels over 50-km Hollow-Core Fiber

Siyu Chen†
*NGIA, School of Optical and Electionic Information*
*Huazhong University of Science and Technology (HUST)*
Wuhan, China
ashley_chen@hust.edu.cn

Zichen Qian†
*NGIA, School of Optical and Electionic Information*
*Huazhong University of Science and Technology (HUST)*
Wuhan, China
zichenqian@hust.edu.cn

Mingming Zhang
(Corresponding Author)
*NGIA, School of Optical and Electionic Information*
*Huazhong University of Science and Technology (HUST)*
Wuhan, China
carlzhang@hust.edu.cn

Huibin Zhou
(Corresponding Author)
*NGIA, School of Optical and Electionic Information*
*Huazhong University of Science and Technology (HUST)*
Wuhan, China
zhouhuibin@hust.edu.cn

Wuyan Wang
*NGIA, School of Optical and Electionic Information*
*Huazhong University of Science and Technology (HUST)*
Wuhan, China
m202573289@hust.edu.cn

Xuchen Hua
*NGIA, School of Optical and Electionic Information*
*Huazhong University of Science and Technology (HUST)*
Wuhan, China
hxctj@hust.edu.cn

Yunlong Li
*NGIA, School of Optical and Electionic Information*
*Huazhong University of Science and Technology (HUST)*
Wuhan, China
liyunlong@hust.edu.cn

Boran Duan
*NGIA, School of Optical and Electionic Information*
*Huazhong University of Science and Technology (HUST)*
Wuhan, China
m202673677@hust.edu.cn

Yuan Gao
*State Key Laboratory of Optical Fibre & Cable Manufacture Technology*
*Yangtze Optical Fibre and Cable Joint Stock Limited Company (YOFC)*
Wuhan, China
gaoyuan_5001643@yofc.com

Peng Li
*State Key Laboratory of Optical Fibre & Cable Manufacture Technology*
*Yangtze Optical Fibre and Cable Joint Stock Limited Company (YOFC)*
Wuhan, China
lipeng@yofc.com

Lei Zhang
*State Key Laboratory of Optical Fibre & Cable Manufacture Technology*
*Yangtze Optical Fibre and Cable Joint Stock Limited Company (YOFC)*
Wuhan, China
zhanglei@yofc.com

An Pan
*Wuhan ANPI Optoelectronics Company, Ltd*
Wuhan, China
panan@anpi.ltd

Cheng Zeng
*NGIA, School of Optical and Electionic Information*
*Huazhong University of Science and Technology (HUST)*
Wuhan, China
zengchengwuli@hust.edu.cn

Yingbin Xing
*NGIA, School of Optical and Electionic Information*
*Huazhong University of Science and Technology (HUST)*
Wuhan, China
xingyb@hust.edu.cn

Hao Wu
*NGIA, School of Optical and Electionic Information*
*Huazhong University of Science and Technology (HUST)*
Wuhan, China
wuhaoboom@hust.edu.cn

Siqi Yan
*NGIA, School of Optical and Electionic Information*
*Huazhong University of Science and Technology (HUST)*
Wuhan, China
siqya@hust.edu.cn

Jinsong Xia
(Corresponding Author)
*NGIA, School of Optical and Electionic Information*
*Huazhong University of Science and Technology (HUST)*
Wuhan, China
jsxia@hust.edu.cn

Ming Tang
(Corresponding Author)
*NGIA, School of Optical and Electionic Information*
*Huazhong University of Science and Technology (HUST)*
Wuhan, China
tangming@mail.hust.edu.cn

***Abstract*—We experimentally demonstrate simultaneous coherent transmission across 1-μm and C+L bands, carrying 54×125-GBaud PDM-16QAM C+L channels and 2×15-GBaud PDM-QPSK 1-μm channels over 50 km, highlighting 1 μm as a promising additional transmission window while avoiding gas-absorption constraints in C/L-band of HCFs.**

## I. INTRODUCTION

Hollow-core fibers (HCFs) have emerged as a promising transmission medium for future high-capacity optical communication systems owing to their ultra-low loss,

This work is supported by National Natural Science Foundation of China (Nos. 62225110, 62601321, and 62601323), the Hubei Province Technological Innovation Program Project (2025BAA003), Hubei Provincial Natural Science Foundation of China (No. 2025AFB008) and Hubei Optical Fundamental Research Centre.

† These authors contribute equally to this work.

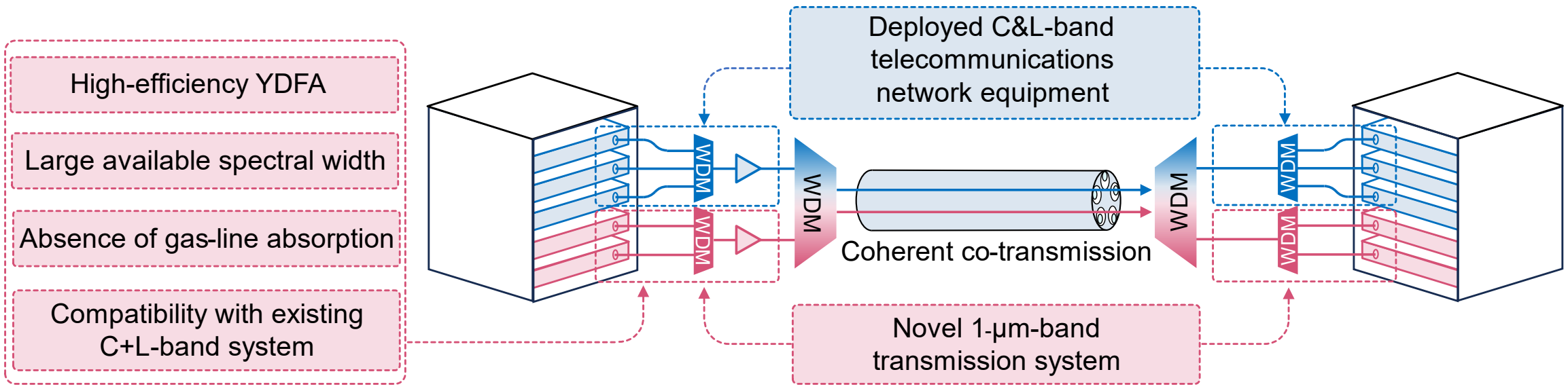

Fig. 1. Characteristics of the 1-μm-band and C+L-band co-transmission system..

extremely low nonlinearity, and reduced propagation latency [1-3]. More importantly, their anti-resonant guidance mechanism provides multiple transmission windows, enabling spectral expansion beyond the conventional telecom bands. In particular, the 1-μm region can form a higher-order anti-resonant transmission window alongside the C/L-band window, offering an attractive additional low-loss spectral region with appropriate fiber optimization [4,5].

The 1-μm band (1020-1080 nm) is particularly attractive for such multi-band transmission. It provides a continuous optical bandwidth of 16.3 THz, more than three times that of the conventional C-band [6,7], and overlaps well with the gain window of high-efficiency ytterbium-doped fiber amplifiers (YDFAs). Furthermore, compared with the C/L bands, the 1-μm window is free from the gas-line absorption (GLA) in HCF link transmission systems caused by major gas molecules in air, making it a compelling candidate for further expanding the usable optical spectrum of HCF systems [8]. These combined advantages make the 1-μm band highly attractive for future ultra-wideband and high-capacity HCF transmission. Meanwhile, currently deployed telecommunications network equipment is still concentrated in the C and L bands. Therefore, a co-transmission system for the 1-μm band and the C+L band over the same link can save fiber resources while remaining compatible with already deployed telecommunications network equipment, providing a low-cost solution for the field deployment of 1-μm band transmission systems.

Previous studies have demonstrated transmission in the 1-μm band, including simultaneous 1-μm and C+L-band transmission over short distances using intensity-modulation/direct-detection (IMDD) systems [9, 10]. However, coherent co-transmission across these widely separated spectral bands has not yet been reported. In this work, we experimentally demonstrate simultaneous coherent transmission in the 1-μm and C+L bands over a 50-km support-tube hollow-core fiber (ST-HCF). The ST-HCF exhibits losses of 0.167-0.254 dB/km in the 1-μm band and 0.091-0.110 dB/km across the C+L bands. Based on this multi-band low-loss link, 2×15-GBaud PDM-QPSK channels at 1 μm are co-transmitted with 54×125-GBaud PDM-16QAM channels across the C+L bands, achieving a net bit rate of 109.86 Gb/s in the 1-μm band after 50-km transmission. Fig. 1 illustrates the characteristics of the novel 1-μm-band and C+L-band co-transmission system.

## II. Experimental Setup

In this experiment, we construct a hybrid transmission link combining a C+L-band WDM coherent optical communication system and a 1-μm-band heterodyne coherent optical communication system. The system setup is shown in

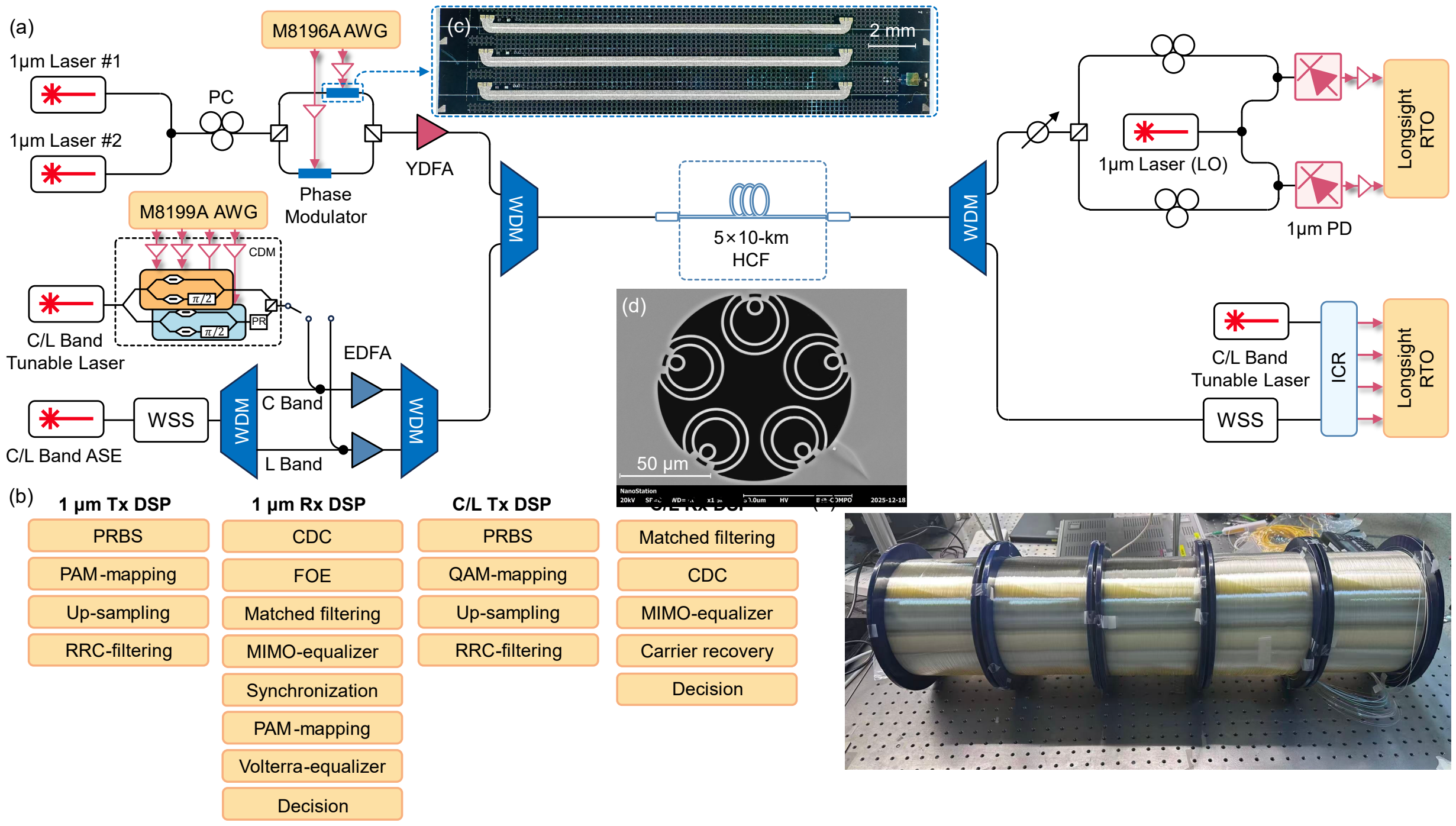

Fig. 2. (a) Experimental setup for the co-transmission of 1-μm and C+L-band communication system; (b) The offline DSP flows;(c) The micrograph of 1-μm-band phase modulator; (d) The cross section of the HCF; (e) The 5×10-km ST-HCF link.

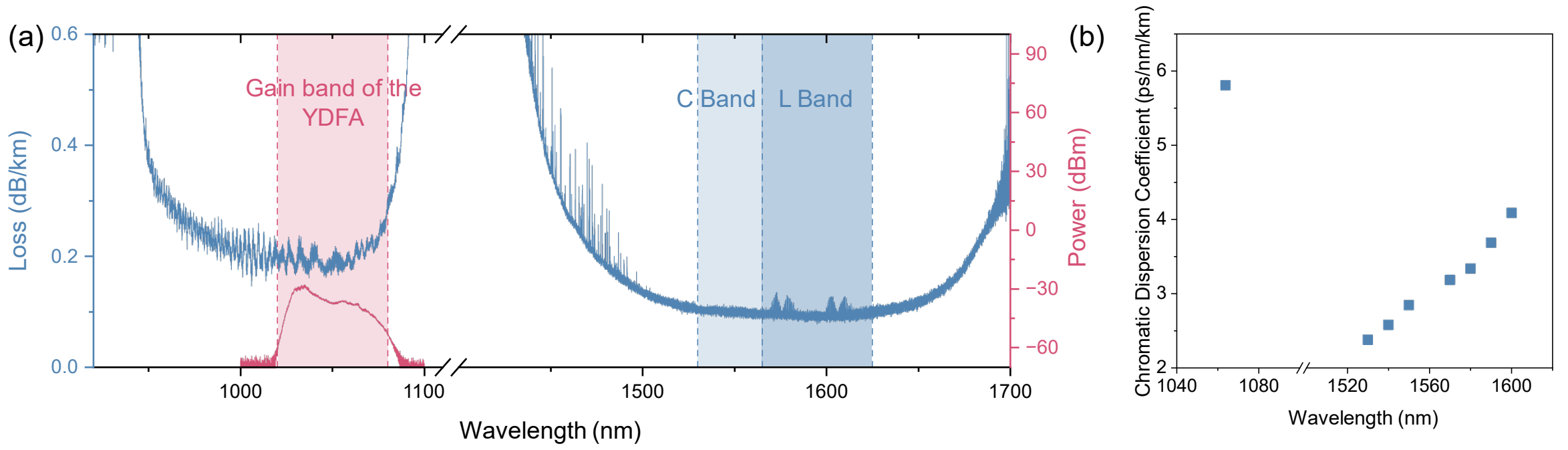

Fig. 3. (a)The measured loss of utilized ST-HCF and the ASE spectrum generated by YDFA; (b) The measured chromatic dispersion coefficient of utilized ST-HCF.

Fig. 2(a), and the DSP strategy at the transmitter and receiver sides for each band is shown in Fig. 2(b).

At the transmitter, the C+L WDM signal consists of the channel under test (CUT) and dummy channels. The dummy channels are generated by a C+L ASE source and shaped by a wave-shaper (WSS) to obtain a flat WDM spectrum, with a per-channel bandwidth of 125 GHz and a channel spacing of 25 GHz. The CUT uses a C/L-band tunable laser, which is fed into a coherent driver modulator (CDM, OVLink-C70-TCL240, 70 GHz). In the transmitter DSP, pseudo−random binary sequence (PRBS) is mapped to 16QAM, and pulse-shaped by a root-raised-cosine (RRC) filter with a roll-off factor of 0.02. The signal is then resampled to match the arbitrary waveform generator (AWG Keysight M8199A, 128 GSa/s, 60 GHz), producing a 125-GBd PDM-16QAM signal. The dummy waves are split into C- and L-band paths by a C/L demultiplexer. The CUT is combined with the corresponding dummy channels via a 3-dB coupler, with a spectral notch reserved for the CUT. The combined signal is amplified by a EDFA and then multiplexed into HCF.

For the 1-μm band, two external-cavity lasers (ECLs) at 1064.25 nm and 1065 nm with a 50-kHz linewidth is combined with a 3-dB coupler. The combined two-wavelength laser source is separated into two orthogonal polarization components by a polarization controller (PC) and a polarization beam splitter (PBS), which are independently modulated by two 1-μm-band phase modulators (ANPI-PM1000, 20 GHz). The modulators are fabricated on a thin-film lithium niobate (TFLN) platform, with a micrograph shown in Fig. 2(c). In the DSP, PRBS is mapped to PAM-4, and then sent to the AWG (Keysight M8196A 92 GSa/s, 32 GHz) to drive the phase modulators, each generating a single-polarization (SP) 15-GBd QPSK signal. Two SP signals are combined into 15-GBd PDM-QPSK signal by a PBC, and amplified by a YDFA, and coupled with the C+L signal through a WDM-MUX before being launched into the ST-HCF. The cross section of the ST−HCF is shown in Fig. 2(d), and Fig. 2(e) displays the 50−km HCF link.

At the receiver, the transmitted signal is separated by a WDM-DEMUX. In the C/L band, the CUT is selected and filtered by another wave-shaper (140 GHz). The signal is detected by an integrated coherent receiver (ICR, OVLink-C70-RCL, 70 GHz) and sampled by a real-time oscilloscope at a sampling rate of 200 GSa/s (RTO, LongSight TS0804A, 80 GHz). The offline DSP includes down sampling, matched filtering, dispersion compensation, CMA equalization, and carrier phase recovery. By sequentially sweeping the spectral notch and the receiver filter, the After demultiplexing, the 1-μm signal passes through a variable optical attenuator (VOA) to control the received optical power (ROP) and is then split into two tributaries by a PBS. The polarization of each one is aligned by a PC. Heterodyne-detection utilizes local oscillators of 17-GHz frequency offset and 1-μm photodetectors (Agilent 86106A, 30 GHz). The RTO captures the detected signals. Offline DSP flows include dispersion compensation, frequency offset estimation, matched filtering, MIMO equalization, PAM-mapping, Volterra equalization, decision, and BER calculation. In a data frame of length 65,536, 1,024 symbols are used as the training sequence to achieve frame synchronization and phase rotation of the signal.

## III. Results

The average loss spectrum of the ST-HCF link is measured using a supercontinuum source and an optical spectrum analyser, shown in Fig. 3(a). Meanwhile, we present the ASE spectrum generated by the YDFA in our system, which characterizes the gain profile of the YDFA. In the C+L band, the loss ranges from 0.091 to 0.110 dB/km. This result reveals that the ST-HCF link exhibits a significantly lower transmission loss compared with conventional silica-core fibers in the 1-μm band. Several GLA features in the L band, where the maximum loss reaches 0.156 dB/km. In contrast, no evident GLA is observed in the 1-μm band. Nevertheless, spectral ripples are present in the 1-μm band, which are attributed to inter-modal interference arising from interference between the fundamental mode and higher-order modes [11]. The chromatic dispersion coefficient of the ST-HCF link is shown in Fig. 3(b).

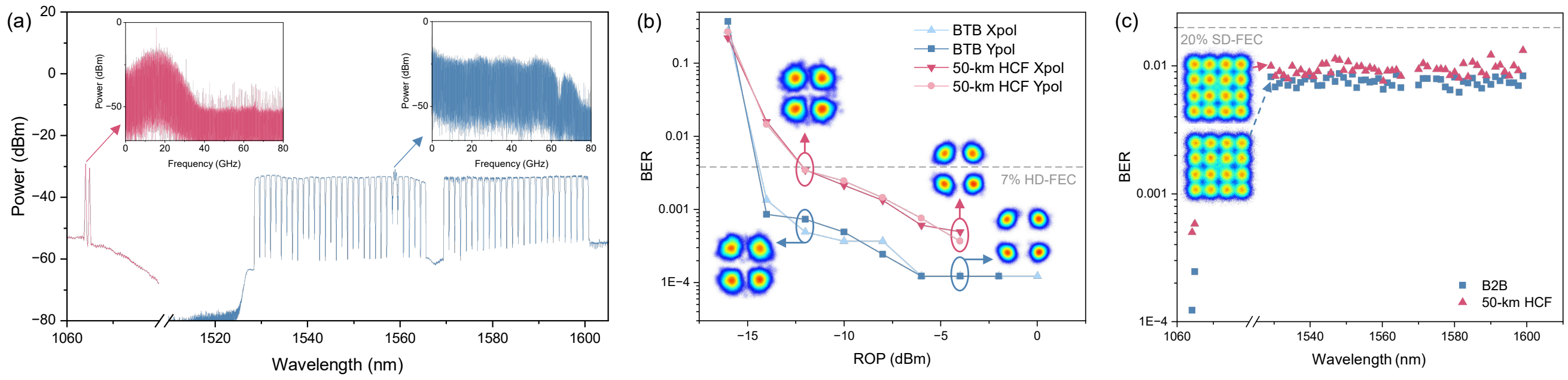

Fig. 4. (a) Optical spectra and RF spectra (insets) of the 1-μm and C+L band signals at the receiver; (b) BER vs. ROP curve of the 15-GBaud PDM-QPSK signal @1064.25 nm; (c) BER performance of the 2 (1-μm) + 54 (C/L) channels.

The co-transmission system, including dummy and CUT signals, is shown in Fig. 4(a). The C and L bands contain 31 and 23 channels, respectively, while the 1-μm band carries two channels centred at 1064.25 and 1065 nm. The corresponding RF spectra of the coherently detected signals are shown in the insets.

Fig. 4(b) compares the receiver sensitivities between back-to-back (BTB) and 50-km HCF transmission in the 1-μm coherent system. The BER as a function of ROP is experimentally measured against the 3.8E-3 HD-FEC threshold. The BERs of both X and Y polarizations are presented to verify the performance of the PDM transmission. The receiver-sensitivity penalty after 50-km transmission is mainly attributed to the relatively strong IMI in the 1-μm band, consistent with the spectral ripples observed in Fig. 3(a).

Fig. 4(c) evaluates the transmission performance at different channels in the C+L band. By sequentially sweeping the wavelength of the CUT, the BERs are measured across 54 channels covering the C+L band against the 2E-2 SD-FEC. The experimental results show that within the measured band, the BER for both the BTB and 50 km HCF schemes remains below the SD-FEC threshold. Meanwhile, owing to the weak GLA, negligible performance degradation is observed for the channels overlapping the GLA. However, this does not imply that such performance can be maintained for longer fiber lengths or with increased gas ingress, where GLA-induced impairment can become more significant. In summary, in the co-transmission system incorporating the 1-μm and C+L bands, the 50-km HCF link exhibits excellent transmission compatibility across a wide spectral range, effectively supporting high-quality transmission for this multi-band system.

## IV. Conclusion

In this paper, we experimentally demonstrate simultaneous coherent transmission in the 1-μm and C+L bands over a 50-km ST-HCF. The ST-HCF exhibits losses of 0.167-0.254 dB/km in the 1-μm band and 0.091-0.110 dB/km in the C+L band, significantly outperforming conventional silica-core fibers at 1-μm. Leveraging this multi-band low-loss transmission, we successfully transmit 2×15-GBaud PDM-QPSK channels in the 1-μm band and 54×125-GBaud PDM-16QAM channels in the C+L band. After 50-km transmission, the net bit rate in the 1-μm band reaches 109.86 Gbps. These results lay the foundation for the low-cost field deployment of novel 1-μm-band coherent optical transmission systems compatible with existing telecommunications network equipment.

## References

[1] Y. Ding, P. Li, G. Chen, J. Chu, A. Jia, L. Zhang, and J. Luo, "Support Tube Hollow-Core Fiber with 0.05 dB/km Attenuation," in *2025 European Conference on Optical Communications (ECOC)*, 2025, pp. 1–4.

[2] M. Petrovich, E. Numkam Fokoua, Y. Chen, H. Sakr, A. I. Adamu, R. Hassan, D. Wu, R. F. Ando, A. Papadimopoulos, S. R. Sandoghchi, G. Jasion, and F. Poletti, "Broadband optical fibre with an attenuation lower than 0.1 decibel per kilometre," *Nature Photonics*, vol. 19, pp. 1203–1208, 2025. DOI: 10.1038/s41566-025-01747-5

[3] L. Feng, H. Liu, K. Chang, J. Luo, A. Zhang, B. Yan, N. Lu, X. Huo, H. Shi, X. Wang, X. Gao, Y. Liu, T. Ma, J. Li, and C. Zhang, "Record Real-Time 2Pbit/s/km Transmission Over 20km Hollow Core Fiber Using Widened C+L 12THz Bandwidth and 1.2Tbit/s Transponder with Symbol Rate of 135GBaud," in *2024 Asia Communications and Photonics Conference (ACP) and International Conference on Information Photonics and Optical Communications (IPOC)*, Beijing, China, 2024, pp. 1–4. DOI: 10.1109/ACPIOPC63121.2024.10809823

[4] H. Sakr, T. D. Bradley, G. T. Jasion, E. Numkam Fokoua, S. R. Sandoghchi, I. A. Davidson, A. Taranta, G. Guerra, W. Shere, Y. Chen, J. R. Hayes, D. J. Richardson, and F. Poletti, "Hollow Core NANFs with Five Nested Tubes and Record Low Loss at 850, 1060, 1300 and 1625nm," in *2021 Optical Fiber Communications Conference and Exhibition (OFC)*, San Francisco, CA, USA, 2021, pp. 1–3.

[5] G. A. Mahdiraji, S. M. A. Mousavi, D. Wu, T. Varghese, N. K. Baddela, M. R. A. Hassan, Z. Kakaei, A. I. Adamu, Z. Zhai, K. Wisniowski, S. B. Gorajoobi, A. Shakiba, M. Rahman, J. Rzegocki, G. Guerra, E. Numkam Fokoua, Y. Chen, G. Jasion, and F. Poletti, "First Dual-Band Hybrid Window Antiresonant HCF with 0.13 dB/km loss at 1 μm and 0.11 dB/km at 1.55 μm," in *OFC 2026*, Th4B.8.

[6] Y. Hong, X. Huang, Y. Jung, H. C. H. Mulvad, H. Sakr, N. Taengnoi, K. R. H. Bottrill, F. Poletti, P. Petropoulos, and D. J. Richardson, "Wideband transmission in the 1-μm band based on a hollow-core fiber and wideband YDFA," in *2023 Optical Fiber Communications Conference and Exhibition (OFC)*, IEEE, 2023, pp. 1–3.

[7] X. Yu, P. Li, L. Dai, J. Li, M. Wu, Y. Zhuo, J. Zhang, O. Xu, S. Fu, and Y. Qin, "T-band 155.7Gbit/s net data-rate DMT signal transmission over 3km NANF-HCF with simple VFFE," in *2025 Optical Fiber Communications Conference and Exhibition (OFC)*, San Francisco, CA, USA, 2025, pp. 1–3.

[8] S. Chen, Y. He, L. Dou, A. P. T. Lau, H. Zhang, Z. Zhai, and Y. Su, "Characteristics and impacts of $CO_2$ absorp tion effects in hollow-core fiber (HCF) transmission systems," in Proc. of ECOC, Copenhagen, Denmark, 2025. DOI: 10.1109/ECOC66593.2025.11263128

[9] N. Yamamoto, Y. Omigawa, K. Akahane, T. Kawanishi and H. Sotobayashi, "Simultaneous triple-waveband, 10-Gbps transmission in 1-μm, C-, and L-wavebands over a 5.4-km long holey fiber," *CLEO/QELS: 2010 Laser Science to Photonic Applications*, San Jose, CA, USA, 2010, pp. 1-2.

[10] D. Cirjulina, T. Salgals, A. Ostrovskis, H. Louchet, M. Koenigsmann, F. Pittala, B. Krüger, L. Zhang, X. Yu, R. Schatz, S. Dahlfort, N. Grossard, R. Jahn, K. Yamaguchi, M. Gruen, V. Bobrovs, X. Pang, and O. Ozolins, "High-Speed Short-Reach Optical Transmission in the 1-μm Band Using a 1064 nm Lithium Niobate MZM," *Optics Letters*, Sep. 2026. DOI: 10.1364/OL.608316

[11] P. Li, G. Chen, A. Jia, H. Li, J. Chu, and Y. Liu, "Low Intermodal Interference and Low Loss Hollow Core Fibers," in *2026 Optical Fiber Communications Conference and Exhibition (OFC)*, 2026, pp. 1–3.